\PassOptionsToPackage{unicode}{hyperref}
\PassOptionsToPackage{hyphens}{url}
\documentclass[
]{article}
\usepackage{amsmath,amssymb}
\usepackage{lmodern}
\usepackage{ifxetex,ifluatex}
\ifnum 0\ifxetex 1\fi\ifluatex 1\fi=0 % if pdftex
  \usepackage[T1]{fontenc}
  \usepackage[utf8]{inputenc}
  \usepackage{textcomp} % provide euro and other symbols
\else % if luatex or xetex
  \usepackage{unicode-math}
  \defaultfontfeatures{Scale=MatchLowercase}
  \defaultfontfeatures[\rmfamily]{Ligatures=TeX,Scale=1}
\fi
\IfFileExists{upquote.sty}{\usepackage{upquote}}{}
\IfFileExists{microtype.sty}{% use microtype if available
  \usepackage[]{microtype}
  \UseMicrotypeSet[protrusion]{basicmath} % disable protrusion for tt fonts
}{}
\makeatletter
\@ifundefined{KOMAClassName}{% if non-KOMA class
  \IfFileExists{parskip.sty}{%
    \usepackage{parskip}
  }{% else
    \setlength{\parindent}{0pt}
    \setlength{\parskip}{6pt plus 2pt minus 1pt}}
}{% if KOMA class
  \KOMAoptions{parskip=half}}
\makeatother
\usepackage{xcolor}
\IfFileExists{xurl.sty}{\usepackage{xurl}}{} % add URL line breaks if available
\IfFileExists{bookmark.sty}{\usepackage{bookmark}}{\usepackage{hyperref}}
\hypersetup{
  pdftitle={How to build an Open Science Monitor based on publications? A French perspective},
  pdfkeywords={open access, open science, open data, open source},
  hidelinks,
  pdfcreator={LaTeX via pandoc}}
\usepackage[left=3cm, right=3cm, top=3cm, bottom=3cm]{geometry}
\ifluatex
  \usepackage{selnolig}  % disable illegal ligatures
\fi
\newlength{\cslhangindent}
\newlength{\csllabelwidth}
 {% don't indent paragraphs
  \setlength{\parindent}{0pt}
  \ifodd #1 \everypar{\setlength{\hangindent}{\cslhangindent}}\ignorespaces\fi
  \ifnum #2 > 0
  \setlength{\parskip}{#2\baselineskip}
  \fi
 }%
 {}
\usepackage{calc}

\newenvironment{cslreferences}%
  {\setlength{\parindent}{0pt}%
  \everypar{\setlength{\hangindent}{\cslhangindent}}\ignorespaces}%
  {\par}

\title{How to build an Open Science Monitor based on publications? A
French perspective}
\usepackage{authblk}
\author[%
  2%
  ]{%
  Laetitia Bracco%
}
\author[%
  1%
  ]{%
  Eric Jeangirard%
}
\author[%
  1%
  ]{%
  Anne L'Hôte%
}
\author[%
  3%
  ]{%
  Laurent Romary%
}
\affil[1]{French Ministry of Higher Education and Research, Paris,
France}
\affil[2]{University of Lorraine, France}
\affil[3]{INRIA, France}
\date{December 2024}

\makeatletter
\def\@maketitle{%
  \newpage \null \vskip 2em
  \begin {center}%
    \let \footnote \thanks
         {\LARGE \@title \par}%
         \vskip 1.5em%
                {\large \lineskip .5em%
                  \begin {tabular}[t]{c}%
                    \@author
                  \end {tabular}\par}%
                                                \vskip 1em{\large \@date}%
  \end {center}%
  \par
  \vskip 1.5em}
\makeatother

\begin{document}
\maketitle
\begin{abstract}
Many countries and institutions are striving to develop tools to monitor
their open science policies. Since 2018, with the launch of its National
Plan for Open Science, France has been progressively implementing a
monitoring framework for its public policy, relying exclusively on
reliable, open, and controlled data.

Currently, this monitoring focuses on research outputs, particularly
publications, as well as theses and clinical trials. Publications serve
as a basis for analyzing other dimensions, including research data,
code, and software. The metadata associated with publications is
therefore particularly valuable, but the methodology for leveraging it
raises several challenges.

Here, we briefly outline how we have used this metadata to construct the
French Open Science Monitor.
\end{abstract}


\hypertarget{requirements}{%
\section{1. Requirements}\label{requirements}}

\hypertarget{data-to-gather}{%
\subsection{1.1 Data to gather}\label{data-to-gather}}

The starting point for these analyses is a corpus of publications.
Defining the appropriate target scope is essential to provide relevant
insights. Details about the metadata required for this corpus are
provided in Section 2. In summary, describing this corpus with a PID
(Persistent Identifier) and associated metadata is crucial. The default
primary PID should be the Crossref DOI. Other PIDs can be used, but the
methodology and code must be adapted accordingly, particularly for Open
Access (OA) status discovery.

Additional metadata are also required. A normalized scientific field
classification enables the creation of KPIs by scientific domain. If
such metadata are unavailable, they can be inferred using machine
learning models. OpenAlex also provides computed metadata that can be
leveraged. Metadata such as publishers, repositories, journals,
affiliations, and publication types can support further analyses, but
they must be normalized to ensure that the insights derived are
meaningful.

Depending on the context, the corpus can be extracted from CRIS systems,
global databases (such as OpenAlex), or even custom-built. The French
Open Science Monitor (OSM) opted for the latter approach, combining open
data from Crossref, PubMed, open repositories, web crawling, and
bottom-up data contributions from French institutions that wish to
develop their own OSM. For global data sources, country affiliation is
determined based on harvested (or crawled) raw affiliation strings using
the affiliation matcher detailed in (L'Hôte and Jeangirard 2021). The
code and a Docker image are available here:
\url{https://github.com/dataesr/affiliation-matcher}.

For Open Access publication KPIs, an OA status discovery tool is
required. By default, Unpaywall provides information for Crossref DOIs.

For indicators related to datasets and software, in addition to the
metadata corpus, the full texts of publications in PDF format are also
required. Text and Data Mining (TDM) techniques can then be applied to
compute the KPIs.

\textbf{Warning}: To achieve the best possible results, it is essential
to download as many full-text publications as possible. In the European
context, this is feasible under the framework of the European directive
allowing text and data mining for research purposes\footnote{Directive
  (EU) 2019/790 of the European Parliament and of the Council of 17
  April 2019 on copyright and related rights in the Digital Single
  Market and amending Directives 96/9/EC and 2001/29/EC
  \url{https://eur-lex.europa.eu/legal-content/EN/TXT/?uri=CELEX:32019L0790}}.
It is necessary to have lawful access to the downloaded content if it is
not already openly accessible, for instance, via a subscription. Outside
the European Union, the legal framework must be carefully reviewed. This
note does not address the context beyond Europe.

\hypertarget{open-source-software-used}{%
\subsection{1.2 Open source software
used}\label{open-source-software-used}}

The French Monitor code is freely available under open license (MIT
License). It is modular as detailed in the infrastructure section in
(Bracco et al. 2022) (for OA to publications) and in (Bassinet et al.
2023) (for datasets and software). However, it is closely linked to our
data acquisition pipeline (web crawling, bottom data collection from
French institutions, extension to cover French specific PID (HAL)) and
to the data architecture we built on the OVH public cloud - S3 Object
Storage used by the Ministry. Also, Python has been used. If it is to be
implemented in another country, parts of the code will have to be
rewritten to match the local requirements. The core monitor code also
relies on other free and open source services / software:

\begin{itemize}
\item
  Open access discovery tool (as explained above): Unpaywall The premium
  service of Unpaywall is used to get a quarterly full snapshot of the
  database. These snapshots are used to historicize the OA status of
  each publication, useful to analyse the OA dynamics.
\item
  A Text and Data Mining (TDM) tool to detect research datasets mentions
  from the full-text. We use
  \href{https://github.com/kermitt2/datastet}{DataStet} from the
  \href{https://hub.docker.com/r/grobid/datastet/tags}{Docker image
  0.8.0}
\item
  A Text and Data Mining (TDM) tool to detect code and software mentions
  from the full-text. We use
  \href{https://github.com/softcite/software-mentions}{Softcite} from
  the
  \href{https://hub.docker.com/r/grobid/software-mentions/tags}{Docker
  image 0.8.0}
\item
  A smart scholarly PDF parsing tool to structure metadata and the
  full-text content from a PDF. We use
  \href{https://github.com/kermitt2/grobid}{GROBID} from the
  \href{https://hub.docker.com/r/grobid/grobid/tags}{Docker image
  0.8.0}. Later versions (from 0.8.1) include fixes on grant ids
  detection that can be very relevant.
\end{itemize}

We also developed extra modules that glues together and orchestrates all
the previous tools:

\begin{itemize}
\item
  The module
  \href{https://github.com/dataesr/bso3-analyse-publications/blob/main/application/server/main/tasks.py\#L99}{bso3-analyse-publications}
  implements in Python the whole
  \href{https://github.com/dataesr/bso3-analyse-publications/blob/main/application/server/main/tasks.py\#L99}{TDM
  pipeline}, and also the analysis
  \href{https://github.com/dataesr/bso3-analyse-publications/blob/main/application/server/main/tasks.py\#L176}{funnel
  analysis} at the document level of the resulting outputs to get
  publication-wise KPIs.
\item
  The module
  \href{https://github.com/dataesr/bso-publications}{bso-publications}
  implements in python an extract-transform-load process and stores the
  final results, at the publication level, in an Elasticsearch index.
\end{itemize}

\hypertarget{computation-power-consumed}{%
\subsection{1.3 Computation power
consumed}\label{computation-power-consumed}}

We deploy our code on a public cloud infrastructure, by OVH. We use
their managed Kubernetes service, using multiple nodes (servers):

\begin{itemize}
\item
  1 large server (16 CPU 240Go RAM) is used to host the metadata Mongo
  databases (Unpaywall snapshots, French corpus metadata) and run most
  of the OA KPIs calculations.
\item
  6 smaller servers are used for data acquisition (crawling, parsing,
  harvesting) and enrichment (language detection, discipline inference,
  tasks monitoring \ldots) (4 CPU 15Go RAM and 2 CPU 7Go RAM).
\item
  1 medium server (4 CPU 60Go RAM) used to harvest the PDF.
\item
  5 servers (32 CPU 120Go RAM) to run the TDM analysis.
\item
  3 servers (with redundancy) (16 CPU 60Go RAM) to host the
  Elasticsearch resulting indices.
\item
  6 servers (with redundancy + staging/prod) (2 CPU 7Go) to host the
  website.
\end{itemize}

This infrastructure is also used for other projects, so 100\% cannot be
affected to the OS monitoring (in particular the Elasticsearch and
websites hosting part). However, the PDF harvesting and TDM analysis are
a very specific need. For this specific need, around 20k euros were
spent to analyse 700k PDFs in one month. The rest of the infrastructure
is around 70k a year, but is not specific to OS monitoring. Relying on
other services (like OpenAlex) could help reduce the costs.

\hypertarget{team-and-human-resources}{%
\subsection{1.4 Team and human
resources}\label{team-and-human-resources}}

A deep understanding of the Open science / scholarly communications area
is key to make it happen. Random software engineers do not have this
knowledge and it may take time for them to understand what is at stake
for the monitoring to be relevant. Building the whole pipeline can be
implemented with about 2 FTE for 6 months. An extra FTE (project manager
or so) can be needed to make sure software developments are inline with
the project goals. Maintenance costs are lower, about 0.5 FTE a year.
However, things evoles fast and new features (new objects to monitor,
new types of analysis \ldots) are generally necessary so a maintenance
only scenario is not very likely to happen.

\hypertarget{a-few-methodological-considerations}{%
\section{2. A few methodological
considerations}\label{a-few-methodological-considerations}}

\hypertarget{corpus-creation}{%
\subsection{2.1 Corpus creation}\label{corpus-creation}}

\hypertarget{defining-a-perimeter}{%
\subsubsection{2.1.1 Defining a perimeter}\label{defining-a-perimeter}}

Research outputs can be indexed in large databases (Crossref, Datacite,
OpenAlex), or not. It can be necessary to put in place specific
harvesters to get metadata from other places (disciplinary or
institutional repositories for example).

Making sure these extra data are correctly ingested with the data coming
from the large database is a challenge (data format, no duplicates
\ldots). An option could be to make those extra research outputs fit
into one of these databases: adding DOI for example, or asking OpenAlex
to harvest extra repositories.

Also, about the distinction between types of publication, for example,
the distinction with professional articles: it all depends on the
available data, of course, but also the main point to define the
perimeter is to know upfront what is the goal of the monitoring. If it
is to steer and analyze the impact of a public policy, then the
perimeter has to be in line with the public policy itself. That may be
different from one national/institutional situation to another. That is
also why, in France, we propose a national monitoring, but also
``local'' monitoring in which the perimeter can be customed by the
users.

\hypertarget{handling-duplicates}{%
\subsubsection{2.1.2 Handling duplicates}\label{handling-duplicates}}

Detecting duplicates can be challenging. For publications with a
Crossref DOI, this PID is in general enough, even though, for some
cases, preprints and published versions sometimes have a Crossref DOI
each and could be considered as only one research output.

For works from Datacite, that can be much more complicated. The Datacite
API has a field ``relatedIdentifiers'' that can be used to detect
duplicates.

In the example from
\href{https://openalex.org/works?page=1\&filter=authorships.author.id\%3Aa5044430271,display_name.search\%3Aconverging}{OpenAlex}
one can see from the datacite API
https://api.datacite.org/dois/10.5281/zenodo.8042997 that this DOI has
several versions. However that may not be enough and some heuristics
based on title, authors \ldots{} or even more complicated processes
(machine learning etc) could be used to detect duplicates. The same goes
for publications from other systems (open repositories for example) that
may contain a lot of duplicates.

Also, during the deduplication phase, duplicate records may have to be
merged into one single record. The list of affiliations, keywords, and
authors can be different. Merging strategies have to be settled. In the
French case, for key elements, Crossref data have the top priority,
followed by the other sources (web scraping, PubMed, HAL, \ldots). For
some metadata that are of type list, it is possible to enrich the final
metadata with all the metadata coming from each source (affiliations,
keywords for example).

\hypertarget{author-disambiguation-and-affiliations}{%
\subsubsection{2.1.3 Author disambiguation and
affiliations}\label{author-disambiguation-and-affiliations}}

Grobid can help to get authors' PID (like ORCID) or raw affiliation
strings and PID (like ROR) when they are present in the full text.
However other sources have to be considered to get better coverage.

Affiliations can be harvested from the landing pages in many cases (from
HTML Highwire header or from HTML parsing). That is one the techniques
used by OpenAlex. Aligning those raw affiliation strings to ROR is
another step where heuristics or machine learning can be helpful (see
(L'Hôte and Jeangirard 2021)). The
\href{https://works-magnet.esr.gouv.fr/}{Works-magnet
https://works-magnet.esr.gouv.fr/} (see (Jeangirard, Bracco, and L'Hôte
2024)) is a tool designed to help improve this matching done
automatically in OpenAlex. Author disambiguation is not an easy task
either as the majority of authors have no ORCID at all. OpenAlex uses
clustering techniques
https://github.com/ourresearch/openalex-name-disambiguation/tree/main/V3

In the French case, we benefit from a large registry of persons
(https://www.idref.fr/) maintained by ABES. This registry has quite a
good coverage of the researchers working in France. Again, clustering
techniques can be used to disambiguate author name with a PID, but a
good person registry helps. These techniques are used in the French
research portal scanR (https://scanr.enseignementsup-recherche.gouv.fr/)
to add PID (namely idref ID) to the authors of publications.

\hypertarget{corpus-enrichment}{%
\subsection{2.2 Corpus enrichment}\label{corpus-enrichment}}

Once the publication corpus is defined, many enrichment are necessary to
add dimensions to analyse. The use of machine-learning techniques can
help a lot, but additional care is required to verify the results (see
(Jeangirard 2022)).

\hypertarget{discipline-classification}{%
\subsubsection{2.2.1 Discipline
classification}\label{discipline-classification}}

Different algorithms based on title, ISSN, authors etc \ldots{} can be
used to classify publications. OpenAlex implements some. The French Open
Science Monitoring uses its own to classify publications into 10
macro-categories (cf (Jeangirard 2019)).

\hypertarget{open-access-discovery}{%
\subsubsection{2.2.2 Open Access
discovery}\label{open-access-discovery}}

The Open Access status information is generally not retrieved with
Grobid (publication parser), but instead using information from the
landing page (HTML parser) or from repositories. Unpaywall (and OpenAlex
now) already implements that logic, at least for all Crossref DOI. For
other publications (with no Crossref DOI), open access discovery is not
that easy and depends on local specificities. In the future, we could
expect that, if those publications are in OpenAlex, they also benefit
from a better open access discovery service just like with Unpaywall.
Also, it is important to note that the open access status is not a fixed
metadata (contrary to the title or the list of authors for instance). It
can evolve over time. More details on this aspect are given in (Bracco
et al. 2022).

\hypertarget{open-access-types}{%
\subsubsection{2.2.3 Open Access types}\label{open-access-types}}

In the French Open Science Monitor, we analyze in different ways the
type of open access. In particular,

\begin{itemize}
\item
  is it opened via the publisher or via a repository (or both?) -
  regardless of the license
\item
  if it is opened via the publisher, what kind of business model is it
  used?
\item
  if it is opened via the publisher, is there any proper license, and
  which one?
\end{itemize}

More details are described in (Bracco et al. 2022).

\hypertarget{apc-estimation}{%
\subsubsection{2.2.4 APC estimation}\label{apc-estimation}}

Estimating Article Processing Charge (APC) is not an easy task. In the
French OSM, we implemented an article level estimation, based on the
data from DOAJ and OpenAPC (cf (Bracco et al. 2022)). In particular, if
an article, not present in OpenAPC, is published in a journal that is
sufficiently represented in OpenAPC (for the same year of publication),
an APC amount is estimated based on the average APC paid for articles
published in that journal in the same year. OpenAlex provides also some
insights, based in particular on listed APC. However, two main
difficulties are difficult to overcome. First, when an article is in
collaboration, it is very difficult to know which institution has
actually paid. The institution of the corresponding author is probably a
good guess, but the corresponding author metadata is hard to get.
Another difficulty comes from the transformative agreements, where APC
are negotiated at a larger scale. The marginal APC per article is then
harder to estimate.

\hypertarget{dataset-and-software-mention-detection}{%
\subsubsection{2.3 Dataset and software mention
detection}\label{dataset-and-software-mention-detection}}

Research datasets, software and code are research outputs difficult to
monitor. We explored a very generic technique to analyse those objects:
this technique can be applied to any country and discipline, as long as
a publication corpus is available. Publications' full-texts are analyzed
with deep learning models to detect all the mentions of data / datasets,
and also all the mentions of software and code. Those mentions are then
caracterized in context, that means they are classified according to the
type of mention. In the model we use, each mention can be a mention of
usage, creation, or sharing. Once all the mentions detected are
caracterized, document-level indicators are computed, to calculate
whether a publication does use / create / share data or software.
Eventually, the French OSM computes national-level indiators, with the
percentage of publication that mentions sharing data (/ software)
amongst the publications that indicates using and producing data (/
software), see (Bassinet et al. 2023).

The detection models (Softcite and Datastet) can be improved for sure.
The \href{https://works-magnet.esr.gouv.fr/}{works-magnet
https://works-magnet.esr.gouv.fr/} provides a module to explore and
correct the mentions detected in the French corpus. We hope to collect
enough good quality curated data to build an extended training dataset
and then more accurate machine learning models for dataset and software
detection. We are convinced setting in place quick human feedback loop
with increase the accuracy of the detection models. Works-magnet like
tools enable that kind of interactions and could be integrated in
different pipelines, like on manuscript deposit platforms.

\hypertarget{a-few-advices-and-impact}{%
\section{3. A few advices and impact}\label{a-few-advices-and-impact}}

Having an objective and quantitative tool for monitoring, easy to plug,
which makes it easier for funders and institutions to communicate and
steer their open science policy. Having a tool that can be easily
adapted to different contexts like funders or institutions is then a key
component. However, a tool remains a tool and is only complementary to
the local policies and mandates.

Some tips also can help in building a reliable and effective Open
Science Monitoring.

\begin{itemize}
\item
  Clean up OpenAlex affiliations to obtain a reliable corpus of
  publications (the
  \href{https://works-magnet.dataesr.ovh/}{works-magnet} can help).
\item
  Maintaining its own infrastructure is costly (money and HR). A call to
  open infrastructure could help reduce the costs and invest in a shared
  infrastructure / methodology.
\item
  If a national dashboard cannot be created, it is already possible to
  obtain useful indicators on open access publications directly via COKI
  (see (Diprose 2023)), as it is based on OpenAlex data.
\item
  In the event of legal and/or economic difficulties in accessing
  non-open access content, a downgraded version of the result based
  solely on open access full text and under CC licence is possible.
\end{itemize}

\hypertarget{references}{%
\section*{References}\label{references}}
\addcontentsline{toc}{section}{References}

\hypertarget{refs}{}
\begin{cslreferences}
\leavevmode\hypertarget{ref-bassinet:hal-04121339}{}%
Bassinet, Aricia, Laetitia Bracco, Anne L'Hôte, Eric Jeangirard, Patrice
Lopez, and Laurent Romary. 2023. ``Large-scale Machine-Learning analysis
of scientific PDF for monitoring the production and the openness of
research data and software in France.''
\url{https://hal.science/hal-04121339}.

\leavevmode\hypertarget{ref-bracco:hal-03651518}{}%
Bracco, Laetitia, Anne L'Hôte, Eric Jeangirard, and Didier Torny. 2022.
``Extending the open monitoring of open science.''
\url{https://hal.science/hal-03651518}.

\leavevmode\hypertarget{ref-coki}{}%
Diprose, Hosking, J. P. 2023. ``User-Friendly Dashboard for Tracking
Global Open Access Performance.'' \emph{The Journal of Electronic
Publishing}. \url{https://doi.org/10.3998/jep.3398}.

\leavevmode\hypertarget{ref-jeangirard_monitoring_2019}{}%
Jeangirard, E. 2019. ``Monitoring Open Access at a National Level:
French Case Study.'' In \emph{ELPUB 2019 23d International Conference on
Electronic Publishing}. OpenEdition Press.
\url{https://doi.org/10.4000/proceedings.elpub.2019.20}.

\leavevmode\hypertarget{ref-jeangirard:hal-04598201}{}%
Jeangirard, Eric, Laetitia Bracco, and Anne L'Hôte. 2024. ``Works-magnet
: aucune de perdue, 10 000 de retrouvées.'' Abes; Journées Abes 2024.
\url{https://doi.org/10.5281/zenodo.11471247}.

\leavevmode\hypertarget{ref-jeangirard:hal-03819060}{}%
Jeangirard, Éric. 2022. ``L'utilisation de l'apprentissage automatique
dans le Baromètre de la science ouverte~: une façon de réconcilier
bibliométrie et science ouverte~?'' \emph{Arabesques}, no. 107
(September): 10--11. \url{https://doi.org/10.35562/arabesques.3084}.

\leavevmode\hypertarget{ref-lhote:hal-03365806}{}%
L'Hôte, Anne, and Eric Jeangirard. 2021. ``Using Elasticsearch for
entity recognition in affiliation disambiguation.''
\url{https://hal.science/hal-03365806}.
\end{cslreferences}

\end{document}